\documentstyle[aps,prl,tighten,psfig]{revtex}
\begin{document}
%

\title{
On depolarisation level shift in spherical QD
}

\author{Slava V. Rotkin}

\address{Ioffe Institute, 26 Polytehnicheskaya, St.Petersburg
194021, Russia;\\
Beckman Institute, UIUC, 405 N.Mathews, Urbana, IL 61801, USA. \\
{\ E-mail: \ \ rotkin@uiuc.edu}
}


\maketitle

\begin{abstract}
A giant level shift, resulted from the interaction
of an electron in a spherical quantum dot
with zero--point oscillations of confined modes of the electric
field, is divulged. The energy correction depends on the dot
radius. This size scaling of the depolarisation effect is
computed semiclassically. A change of the optical properties of
the matrix surrounding the dot provides a method to study the
shift experimentally.
\end{abstract}

%
%

\section{Introduction}
\label{sec:intro}

A complete quantum dot (QD) theory, taking into account all the
sophisticated physics of the object, is still a challenge for a
theorist. The main reason is that the scale of the calculation is
much larger than atomic one (that complicates {\it ab initio}
techniques). The same time the number of particles is small to
use solid state approximations in full extent. For example, an
one--electron picture of a quantum confinement potential, arising
from the conduction band discontinuity on the QD boundary, does
not always yield accurate electron levels.

In the paper we put forward a model to inspect the
electrodynamical correction to the one--electron energy in a
spherical QD. It was shown\cite{lsc60} that the similar
correction occurs to be significant for a "natural quantum dot"
C$_{60}$. A {\em depolarisation} level
shift due to the interaction with an electromagnetic field is
not negligible, as it might be thought, when taking into account
{\em localized} electromagnetic modes.
We present the scaling analysis of some different mechanisms for
the level shift (LS) and propose a possible experimental
manifestation of the depolarisation effect.

In order to appraise the LS a simple {\em spherical} QD
model in frame of an {\em effective mass} approximation was
applied. How is our result sensitive to the model used? The size
scaling of the depolarisation shift preserves, being dependent
mainly on a corresponding density of states of the field, while
the prefactor might be smaller within other approach, though it
is not easy to evaluate explicitly.

The group of full rotations, SO(3), was chosen to label the
one--electron states. It is possible to perform an analytic
quantum--mechanical calculation of the RPA response within the
spherical model\cite{ftt95}. The massive peak of a collective
excitation is known to show up in the spectrum, resulting from
the fast coherent oscillation of the total electron density of
the valence states. Thus, within our model the electron--electron
interaction is dealt with selfconsistently. Of course, the number
of valence electrons involved in the collective motion has not to
be small.  It is believed to fulfill for the typical QD
possessing some hundreds of atoms and even more.

A surface charge density oscillation can be thought as a
confined electric field mode or a multipole surface plasmon.  We
will reflect on the shift of the electron level in the field of
zero--point oscillations of the modes connected with the QD,
which depolarisation effect is billion times stronger than of the
free field zero--point oscillations, so a name "giant LS" is
admitted.

The classical description of the electromagnetic surface modes,
via the dielectric function of the matrix and QD material, gives
the true plasmon state frequencies and will be used below. Once
more, the final result does not depend too much on the
computation approach. Our model sketches out the (many--body)
depolarisation semiclassically, avoiding a lot of the routine
computational intricacy.

The paper proceeds as follows: the brief model description
is next to the introduction. Then, the model will be applied to a
3D--plasmon as well as a free field mode, that will explain the
calculation technique. However, the LSs from these modes are too
small to have an experimental importance. Section
\ref{sec:confined} deals with the confined modes those result in
much larger depolarisation shift. The numerical estimations and
the scaling properties of the LS will be given in respect to a
possible experiment. A brief summary will follow. The calculation
of the QD confined mode frequencies is allocated in Appendix.

\section{Semiclassical theory for energy level shift}
\label{sec:ls}

We have considered semiclassically the LS for an
arbitrary shell object in \cite{hawaii}.
The method follows the one proposed by Migdal\cite{migdal} to
calculate the Lamb shift for a hydrogen--like atom. The
frequency of the zero--point oscillations
of the external field
is much higher than the inverse period of the
electron orbit $\omega_p\gg \pi/\tau$.
Therefore, the adiabatic approximation has to be used and one
divides the fast (field) and slow (electron) variables.
An electron is subjected to short fast deflections from its
original orbit in the high--frequency field of the
electromagnetic wave of the zero--point oscillation.
Then the energy shift is given by the second order perturbation
theory as
\begin{equation}
\delta E= \left< H(r+\delta)-H(r) \right> =
\left< \nabla H\cdot \vec\delta + \frac{1}{2} \nabla^2 H
\; \vec \delta \cdot \vec \delta + \ldots\right>,
\label{ls1}
\end{equation}
where $H(r)$ is the unperturbed Hamiltonian and $H(r+\delta)$ is
the Hamiltonian with account for the random electron deflection
$\delta$. The angle brackets represent the quantum mechanical
average over the fast variables of the field (or, the same, over
the random electron deflections). The perturbed Hamiltonian is
expanded in series on the $\delta$ and a first nonzero
contribution is taken.

The simplest QD Hamiltonian is considered to have only the
rotational correction which is given by:
\begin{equation}
\displaystyle \delta H= \frac{{\hat L}^2}{2mR^2} \left( -2
\frac{\delta}{R}+3\frac{\delta^2}{R^2}+\ldots\right),
\label{ham}
\end{equation}
where $R$ is about the spherical QD radius; $m$ is the electron
mass which is supposed to be constant within the dot; $\hat L$ is
the angular momentum operator. On the averaging, the
first--order term disappears.  So far the LS dependence on the QD
size includes $R^{-4}$ factor besides some power hidden in the
mean square deflection $\overline{\delta^2}$.  The strength of
the electrodynamical interaction changes with this quantity
almost exactly. We will show that the dependence of
$\overline{\delta^2}$ in $R$ is different for different electric
modes (confined and free field).  The giant deflection is
representative for the giant LS and, therefore, the function
$\overline{\delta^2}(R)$ will be studied specifically.

\section{Bulk plasmon contribution to LS}
\label{sec:bulk}

First we consider the bulk 3D--plasmon modes that could shift the
electron level. Nearly self--evidently the bulk plasmon shift is
negligible. The mean square deflection, caused by the 3D mode
(which is not confined at all), decreases with the QD size too
rapidly. The small factor, contained in the 3D LS, comes
essentially from the expression for $\overline{\delta^2}$ which
scales as $1/N$, where $N$ is the number of atoms in the QD. It
will be explained in this section.

Within the semiclassical approach, the deflection of the
electron can be computed with the use of the Newton law:
\begin{equation}
m \partial_t^2 \delta = e {\cal E},
\label{newt}
\end{equation}
here $e$ is the electron charge, ${\cal E}$ is the field strength
due to the zero--point oscillation of some mode, $m$ is the
electron effective mass.
The square of the deflection
$\overline{\delta^2}=e^2/(2m)^2 \int d^Dk \; \overline{{\cal
E}_k^2}/\omega_k^4$ is proportional to the mean square of the
electric field strength.  The dimension of the field, $D$, equals
3.  The field strength, in turn, can be rewritten as the
zero--point oscillation frequency $\overline{{\cal
E}_k^2}=2\pi\hbar \omega_k$ through the quantized field
normalisation.

The scale of the energy is given by the 3D plasmon frequency
$\omega_p=\sqrt{4\pi e^2n_{\rm 3D}/m}$. Note that the
3D plasmon frequency does not depend on the quantum number $\bf
k$ and passes through the integral. Hence, the mean square
deflection contains the total number of states effecting on the
electron level in the QD. The integral is limited above by
$k_{max}\sim 1/R$. In 3D--case it brings the factor $R^{-3}\sim
N^{-1}$ claimed in the beginning of the section.

This result will change for other confined electric modes because
of their different densities of states. This produces the
different $N-$scaling factor for the LS from these modes.

The prefactor of the deflection, for any mode considered here,
depends equally on the square root of the
density of electrons, which is useful to be converted
to $r_s$, a characteristic length via the
following definition: $2\pi r_s^3 n_{\rm 3D}/3=1$.
Then, for 3D plasmon the deflection
reads as:
\begin{equation}
\overline{\delta^2}=a_B^2
\displaystyle \frac{\sqrt{6}}{6^4\pi}
\left( \frac{r_s}{a_B} \right)^{3/2}
\left( \frac{r_s}{R} \right)^3 \propto N^{-1},
\label{3d}
\end{equation}
where the atomic length unit, $a_B=\hbar^2/m e^2\simeq 0.53$\AA,
or the Bohr radius, gives the scale of the deflection
(note that this definition does not include any permittivity
unlike an exciton Bohr radius in semiconductors).

The depolarisation (the ratio of the level shift,
$\delta E$, to the bare
energy, $E^{(o)}$) due to the 3D modes is as follows:
\begin{equation}
\Delta_{\rm 3D}=\displaystyle \frac{\delta E}{E^{(o)}}=
\frac{1}{72\sqrt{6}\pi} \sqrt{\frac{a_B}{r_s}}
\left( \frac{r_s}{R} \right)^5 \propto N^{-5/3}.
\label{ls3d}
\end{equation}
The rude estimation of the prefactor shows that even for the
small QD with $N=100$ the shift is $10^{-6}$ of the bare energy
and will not be resolved because of a number of other different
factors effecting the level position.

To give a complete picture we note that the standard LS due to
the zero--point oscillations of the
free electromagnetic modes of the
vacuum can be written as:
\begin{equation}
\Delta_{\rm vac}=
\frac{6\alpha^3}{\pi} \left( \frac{a_B}{r_s}\right)^2
\left[ \ln \frac{r_s}{\alpha a_B}+ \ln \frac{R}{r_s} \right]
\left( \frac{r_s}{R} \right)^2 \propto N^{-2/3},
\label{lsvac}
\end{equation}
where $\alpha\simeq 1/137$ is the fine structure constant, and
the simple check shows that the logarithmic dependence of the
last term in square brackets on $N$ does not add any extra
to the result and has to be dropped in our case. Though the slope
of the LS in $N$ is much slower than in Eq.(\ref{ls3d}) the
prefactor is tiny ($\sim 10^{-7}$) because of $\alpha^3$.

\section{Depolarisation: confined modes}
\label{sec:confined}

Let us consider the specific behavior of the LS
materialized by the zero--point oscillations of the
confined plasmon modes.
The depolarisation in carbon shell cluster was
shown\cite{lsc60}
to be independent of the cluster size. The mean square
deflection scales also as a zero power of the size
$\overline{\delta^2}\propto N^{0}$. While it is interesting by
itself, the carbon cluster matter will not be considered
in the paper. However, there are confined modes in our
QD problem those enhance the electrodynamical correction to the
electron energy.

With the decrease of the dimension of the field the
plasmon density of states increases. Hence, the mean interaction
of the electron with the plasmon field increases that will be
evident from the scaling of $\overline{\delta^2}$.

Two possible candidates for the confined plasmon modes in the QD
system, those have different densities of states,
are the 2D plasmon and the 0D spherical mode. The former
mode can arise because of some interface possibly grown within
the structure (see inset in the Fig.\ref{fig:lambs}). It might be
a conducting wetting layer, if it is thick enough to confine the
electromagnetic field. The 2D plasmon naturally originates at the
interface between the semiconductor structure and a metal
\cite{ando}. At the boundary of two dielectrics a surface plasmon 
is known to propagate\cite{land8}. Its contribution will be 
discussed elsewhere as being smaller than 2D--plasmon one by a 
factor $10^{-2}$ at least owing to the fast space decay.

The 0D mode is the inherent property of the spherical inclusion
of the foreign material in any matrix. The calculation
of the frequency of this mode is slightly cumbersome (see
Appendix for details). The surface QD mode has the quantum
numbers $L, M$, the angular momentum and its projection on an
axis, instead of 2D wave vector, $\bf k$, for the standard 2D
plasmon modes.  The depolarisation is anomalous large
in the 0D case. It will be seen in this section
from the scaling of $\overline{\delta^2}$ and $\Delta$.

\subsection{2D plasmon}
\label{sec:2d}

The frequency of 2D plasmon is well
known\cite{chaplik} to depend on its 2D wave vector as:
$\omega_k=\sqrt{2\pi e^2 n_{\rm 2D} k/m}$.
We will rewrite the 2D electron
density, as before, in terms of the characteristic length:
$\pi r_s^2 n_{\rm 2D}/2=1$ and perform the
integration over the plasmon states.
Then the mean square deflection can be expressed
as:
\begin{equation}
\overline{\delta^2}=a_B^2 \displaystyle \frac{1}{32}
\left( \frac{r_s}{a_B} \right)^{3/2}
\left( \frac{r_s}{R} \right)^{3/2} \propto N^{-1/2}.
\label{2d}
\end{equation}
The scaling in $N$ has a lower exponent that reflects the
different density of the confined field (plasmon) states.
Substituting $\overline{\delta^2}$ into the Hamiltonian
given by Eq.(\ref{ham}), one
gets the depolarisation as follows:
\begin{equation}
\Delta_{\rm 2D}=
\frac{3}{32} \sqrt{\frac{a_B}{r_s}}
\left( \frac{r_s}{R} \right)^{7/2} \propto N^{-7/6}.
\label{ls2d}
\end{equation}
The shift depends on the inverse size nearly linearly. However,
the prefactor dominates at some moderate size of the QD and
lessens the LS to $10^{-3}$ for $N=100$. The depolarisation is
still to be too small to expect experimental consequences. To be 
precise the result also depends on $w$, the distance between the 
2D electrons and the QD. It is simply included in the 
consideration by multipling Eq.(\ref{ls2d}) by a factor 
$\sqrt{\pi} {\rm Erf}(\sqrt{w/R})/(2\sqrt{w/R})$, and the 
depolarisation declines 4 times at $w/R\sim 10$.

\subsection{QD confined plasmon: Mode of cavity}
\label{sec:qd}

\begin{figure}[h]
\centerline{ \psfig{figure=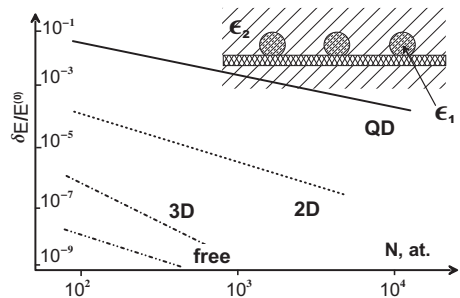,width=6cm}}
\vskip .2cm
\caption{\label{fig:lambs}
The level shift of the QD electron, calculated for 4 different
depolarisation mechanisms.
The giant shift, which results from the zero--point oscillations
of the electric field of the QD surface mode, is shown in full
line. The shifts by the 2D plasmon, the matrix bulk plasmon modes
and the free field are shown as dash lines. The slopes and the
prefactors of these depolarisation shifts are derived
analytically and explained in the text. Inset: The scheme of the
QD structure to model.} \end{figure}

The $\overline{\delta^2}$ considered above the less, the larger
the QD size, that is not the case\cite{lsc60}
for the giant deflection due
to the completely localized modes. The localized modes are the
surface plasmons of the spherical inclusion (with the dielectric
function $\epsilon_1$) in the matrix (with the different
dielectric function $\epsilon_2$).
The frequency of the mode, $\omega_L$, that
we consider, is nearly the frequency of the bulk plasmon in the
matrix,
$\omega_{p2}$, with the weak dependence on the mode angular
momentum (see Appendix).
The electric field of the zero--point oscillation is given
by the formula ${\cal E}_L^2=\pi (L+1/2)\hbar
\omega_L/R^3$. The summation over all states below some critical
value $L_c$ gives the mean square deflection:
\begin{equation}
\overline{\delta^2}=a_B^2 \displaystyle
\frac{\pi}{9\sqrt{6}}
\left( \frac{r_s}{a_B} \right)^{3/2}
\left( \frac{r_s}{R} \right)^3 \left( L_c+\frac{1}{2} \right)^3,
\label{qd}
\end{equation}
where it is natural to limit the summation above the
excitation which wavelength is about
the lattice constant $d$.
We found\cite{contr}
that the $\overline{\delta^2}$ does not depend on the QD
size:
\begin{equation}
\overline{\delta^2}=a_B^2 \displaystyle
\frac{\pi^4}{9\sqrt{6}}
\left( \frac{r_s}{a_B} \right)^{3/2}
\left( \frac{r_s}{d} \right)^3 \propto N^{0}.
\label{qdbis}
\end{equation}
Sequently, the level shift depends on the size as $R^{-2}$
(which comes from Eq.(\ref{ham})):
\begin{equation}
\Delta=\displaystyle \frac{\pi^4}{3\sqrt{6}}
\sqrt{\frac{a_B}{r_s}}
\left( \frac{r_s}{d} \right)^3
\left( \frac{r_s}{R} \right)^2 \propto N^{-2/3}.
\label{lsqd}
\end{equation}
Our estimation shows that the level correction, becoming of the
order of 50\%, plays the important role for the QD
of 100 atoms and smaller.  We collected all studied contributions
to the depolarisation LS and plot them in the log--log scale
versus the QD size in Figure \ref{fig:lambs}.

The depolarisation because of the localized surface QD modes is
large enough to propose an experiment supporting our model.
It is easy to see that $\overline{\delta^2}\sim \omega^{-3}_L$,
whence the LS depends on the mode frequency as well.
Therefore, changing the optical properties of the matrix
surrounding the QD, one shifts the levels. If the bare energy
level, $E^{(o)}$,
lies deep in the potential well, its position is nearly
independent of the well depth which changes along with the matrix
parameters. The deep bare level energy depends only on the well
width $\sim R$.  Hence, keeping the same size of the QD and
covering it with the different materials, one will derive solely
the depolarisation LS, since it is distinguishable from the
standard space quantization LS.

\section{Summary}

The effect of the zero--point oscillations of the free and
confined electromagnetic field on the level of the confined
electron in the spherical QD is reviewed. The depolarisation due
to an interaction with the zero--point oscillations of the field 
(produced by all other valence electrons) shifts up the
bare one-electron state that seems to be a counterpart for the
vertex correction (electron--hole interaction, for example) which
lowers the transition frequency down. It indicates that the
studied effect should be taken into account for a many--body
computation of a QD spectrum.

To the best of our knowledge, the scaling
dependence of the depolarisation level shift for the QDs is
calculated in the first time. The size dependence of the LS
is different for 4 cases considered in the paper. This scaling
reflects that the different densities of states work in different
mechanisms of the depolarisation due to the different 3D, 2D and
0D--dimensional modes of the electric field are involved. Our
model allows a theorist to skip a tedious quantum
electrodynamical calculation but obtain the analytical
selfconsistent estimation for the (many--body) level shift in a
nanoscale system with the strong quantization. The result has not
only a theoretical importance.

Although, the depolarisation decreases with the QD size
in general, the {\bf localized} surface electromagnetic mode (which is
specific to the QD as a void in the matrix material) results
in the giant level shift and is to be possibly resolved
experimentally for the QD made from some hundred atoms. Another
method to detect the effect could be the measurement of a deep
level position in the similar QDs buried by the substances with
the distinct optical characteristics. Then the localized plasmon
frequency changes along with the prefactor of the depolarisation
shift, which could be observed by the optical spectroscopy
of the QD system.

\begin{figure}[h]
\centerline{ \psfig{figure=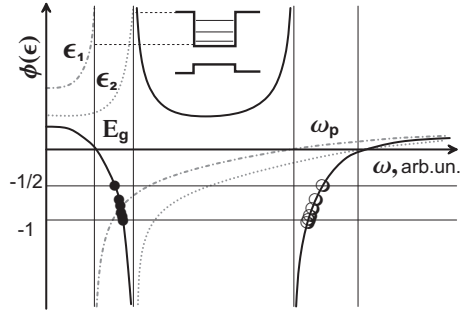,width=6cm}}
\vskip .2cm
\caption{\label{fig:diel}
The graphical solution for the QD plasmon modes with $L\ge 1$.
The solid curve gives the function to solve (see secular equation
in the text). The light dotted curves represent the dielectric
functions of the dot and matrix substances. The full circles are
the solutions for the sphere--like modes; the open circles are
the cavity--like modes (see the text). Inset: The scheme of the 
band structure of a QD system.} 
\end{figure}

\begin{appendix}
\section{Surface QD plasmon modes}
\label{sec:cavity}

The sought--for modes are given in complete spherical harmonics
$P_{L}(r) Y_{L,M}(\Omega)$, where $P_{L}(r)$ is the Legendre
polynomial and $Y_{L,M}$ is the spherical harmonic\cite{abram}.
The electrodynamic solution for the modes of the system
consisting of the spherical particle with the dielectric function
$\epsilon_1(\omega)$ and the surrounding matrix with the
different dielectric function $\epsilon_2(\omega)$ is one of the
roots of the secular equation:
\begin{equation}
\displaystyle \frac{\epsilon_2(\omega)}{\epsilon_1(\omega)}=
-\frac{L}{L+1}.
\label{secul}
\end{equation}
The similar equation gives a mode of an empty void in the
matrix in the limit $\epsilon_1=1$ (in the limit $\epsilon_2=1$
it gives a mode of a sphere in the vacuum). The right hand
side varies from $-1/2$ to $-1$.

Let us suppose the simplest form for the dielectric function
defined by one pole at some frequency, which is about the optical
gap $E_g$, and one zero, which is given by the bulk plasma
frequency of the material (both parameters differ for the
materials 1 and 2).  Then the solutions of the Eq.(\ref{secul})
are easy to find graphically (see. Fig. 2). The function of
$\omega$ standing in the left hand of the expression is plotted
in solid. It has two poles at $E_{g2}$ and $\omega_{p1}$ and two
zeroes at $E_{g1}$ and $\omega_{p2}$. All roots with $L=1, 2..
L_c$ lie in between these two pairs as shown in the figure. The
dependence of the mode frequency on the angular momentum is very
weak.

The lower--frequency modes, shown as black circles, correspond to
the sphere--like plasmons. As lying above the QD gap, the lower
mode 
effectively damps inside the QD. Therefore we
will consider only the upper branch, shown as open circles, which
is similar to the modes of the cavity.  The frequency of the
upper mode is close to the frequency of the bulk plasmon
$\omega_{p2}$ as it is seen from Figure (\ref{fig:diel}). The
mode frequency dependence on its angular momentum is negligible.

The change of any of the 4 parameters, defining $\epsilon_1$ and
$\epsilon_2$, results in the QD mode frequency shift. However,
for the sought--for cavity--like mode, the most important are the
plasma frequencies of the materials. This provides the mechanism
for an experimental observation of the described effect in the
matrix materials with the different $\omega_p$.
It influences on the depolarisation LS, while the 
space quantization of the one--electron level, which is deep 
enough, depends solely on the QD width which has to be kept.

\end{appendix}


\begin{references}
\bibitem{contr}
For the infinitely large sphere, the contraction limit is
fulfilled: $R,N,L\to\infty$, but $R/L=\rm const$, then the
(infinitely large) angular momentum can be related to the
(finite) 2D wave--vector via $L\simeq k R$.
Substituting the maximal wave--number $k_{\rm max}\sim \pi/d$
into this expression we get the maximal angular momentum as
$L_c+1/2\sim \pi R/d$, where $d$ is the lattice constant.
Then the critical angular momentum divided by the
radius becomes some constant $(L_c+\frac{1}{2})/R\sim
\pi/d$.
\end{references}
\end{document}